\documentclass[12pt]{article}
\usepackage{epsf,amsfonts,amssymb,epsfig,amsmath,graphics,slashed}
\usepackage{hyperref,graphicx,subfig}
\begin{document}

\makeatletter
\renewcommand{\theequation}{\thesection.\arabic{equation}}
\@addtoreset{equation}{section}
\makeatother

\baselineskip 18pt

\begin{titlepage}

\vfill

%\begin{flushright}
%Imperial/TP/2011/JG/05\\
%\end{flushright}

%\vfill

\begin{center}
  \baselineskip=16pt
  {\Large\bf Useless Qubits in ``Relativistic Quantum Information''}
 \vskip 1.5cm
    Fay Dowker\\
  \vskip .6cm
     \begin{small}
     \textit{Blackett Laboratory, 
       Imperial College, London, SW7 2AZ, U.K.\\
       and\\
       Perimeter Institute, 39 Caroline St. N., Waterloo, ON N2L 2Y5, Canada
       }
       \end{small}\\*[.6cm]

\end{center}

\begin{center}
\textbf{Abstract}
\end{center}

\begin{quote}
I draw attention to previous work that shows that the observables corresponding to 
relativistic quantum field modes commonly employed in 
papers on ``relativistic quantum information'' cannot be measured by ideal measurements.
\end{quote}

\vfill

\end{titlepage}
\setcounter{equation}{0}

%----------------------------------------------------------------------------------------------------------------------------

What observables can be measured by an ideal measurement 
--  one that leaves an eigenstate 
of the observable in an eigenstate -- in a relativistic quantum field theory (RQFT)? 
The answer to this question is unknown in general but we do know examples of observables 
that cannot be measured if we assume that superluminal signalling (SS) 
is impossible. One example is a Wilson loop in a nonabelian gauge theory \cite{Beckman:2001ck}. 
Another unmeasurable observable, corresponding to 
a one particle state \cite{Sorkin:1993gg}, is however commonly used
to construct qubit Hilbert spaces in papers on ``relativistic quantum information.'' 

I will not rehearse the calculation in \cite{Sorkin:1993gg} but merely emphasise a point made in the paper that
although it is done in Minkowski spacetime, the same calculation can be performed 
without alteration in QFT in any globally hyperbolic spacetime including for example
QFT ``localised'' in a box with reflecting boundary conditions. 
What the calculation shows is that if some specified one particle 
state can be measured by an ideal measurement then two agents inside the box, one to the past of the 
spacelike hypersurface, $\Sigma$, 
on which the observable is measured and one to the future, can signal 
superluminally. 

The same observation can be made about unitary transformations in general: the only unitary transformation 
that is safe to perform on a hypersurface is one which is a product of ``ultralocal'' unitary 
operators at points on the hypersurface. A unitary transformation which is nonlocal on $\Sigma$
generally permits SS. In order to protect relativistic causality 
the ``Local'' in ``Relativistic LOCC'' should be replaced by ``ultralocal'', pertaining to points in
spacetime. 

The problem is that no box is local enough. A RQFT observable such as that corresponding to 
a one particle state is still highly nonlocal within the box. One could say that one simply does 
not allow discussion of  signalling inside the 
laboratory/box of a given external agent, that the lab/box can be made as small as we like 
and is simply defined to be ``local" as far as the analysis goes. This wouldn't 
be convincing: one would hardly 
counter the claims of Beckman et al. by saying, there's no problem in assuming a
Wilson loop can be measured because the agents who could signal superluminally would have to be confined 
inside a proton and we don't allow them.
To belabour the point: it is not enough simply to claim that the ideal measurement of a one particle state 
in a RQFT can model approximately the effect of a real intervention in an actual experiment
 without giving an explanation of why this is valid but other operations 
 in the approximating RQFT (such as those that 
 result in SS) cannot model actual interventions. 
 On the other hand, if  the effect of the actual measurement is 
 not modelled by a projection operator onto the one particle state
 on a spacelike hypersurface, then it would have to be spelled 
 out what the actual measurement {\textit{is}} modelled by in the formalism of the RQFT
 and the consequences for information processing reassessed.
 Beckman et al. describe how local agents can collaborate to measure a Wilson loop 
 by a demolition measurement, for example, but a 
 demolition measurement would not be useful for teleportation or 
 other types of information processing. 
 
The significance of  the ``measures of entanglement" {\textit{etc.}}  calculated
using nonlocal modes states is therefore unclear as far as information processing is 
concerned, no matter how interesting they may be from the point of view of RQFT itself.
The sine qua non of quantum information theory 
is the information held and communicated by external agents and they 
must be able to read it by making measurements on the quantum systems they hold. 
Calculations in RQFT using unmeasurable 
observables have unknown relevance to quantum information.

\section*{Acknowledgments}
I thank Hans Westman for useful discussions. 
This work was supported in part by COST Action MP1006. 
Research at Perimeter Institute for Theoretical Physics is supported in part by the Government of Canada through NSERC and by the Province of Ontario through MRI.

\bibliographystyle{utphys}
\bibliography{refs}

\end{document}